# A Simple 'Range Extender' for Basis Set Extrapolation Methods for MP2 and Coupled Cluster Correlation Energies

Jan M.L. Martin[1, a)]

[1] *Department of Organic Chemistry, Weizmann Institute of Science, 76100 Reḥovot, Israel*

[a)]Corresponding author: gershom@weizmann.ac.il

**Abstract.** We discuss the interrelations between various basis set extrapolation formulas and show that for the nZaPa and aug-cc-pVnZ basis set formulas, for n=4–6 their behavior closely resembles the Petersson $(L+a)^{-3}$ formula with a shift $a$ specific to the basis set family and level of theory. This is functionally equivalent to the Pansini-Varandas extrapolation for large $L$. This naturally leads to a simple way to extend these extrapolations to n=7 and higher. The formula is validated by comparison with newly optimized extrapolation factors for the AV{6,7}Z basis set pairs and literature values for {6,7}ZaPa. For $L \geq 5$, the CCSD extrapolations of both the Schwenke and Varandas type are functionally equivalent to $E(L)=E_\infty+A.(L-0.30)^{-3}$, i.e., $E_\infty=E(L)+[E(L)-E(L-1)]/([(L-0.30)/(L-1.30)]^3-1)$

The correlation energy's very slow convergence with the one-particle basis set has been a major bottleneck for accurate wavefunction ab initio calculations. Explicitly correlated methods (see[1,2] for recent reviews) are an emerging alternative; in conventional orbital-based calculations, basis set extrapolation formulas are widely used.

Schwartz[3,4] was the first to show that the second-order correlation energy of helium-like atoms converges as $A.L^{-3}+B.L^{-5}$, where L is the largest angular momentum being considered. This work was generalized by Hill[5] and Carroll[6] and by Kutzelnigg and Morgan.[7] It was realized early on[8,9] that this could be turned into an extrapolation formula rather than a pessimistic error estimate, and since there has been a proliferation of such formulas based on the partial-wave expansion. To single out just a few: Martin[9] considered (see also, e.g., Ref.[10]):

$$E(L) = E_\infty + A.L^{-\alpha} \qquad (1)$$

where α is an adjustable exponent that can either be determined by 3-point extrapolation[9] or (for 2-point extrapolation) set to a fixed value determined by fitting to a training set.[10,11]

Feller[12] proposed a simple 3-point geometric extrapolation on purely empirical grounds. It almost invariably represents a lower limit to dissociation energies: [13]

$$E(L) = E_\infty + A.e^{-B.L} \qquad (2)$$

The simple Helgaker two-point formula[14] (eq. (1) with fixed α=3) is of course the most widely used. It typically represents an upper limit to dissociation energies:[13]

$$E(L) = E_\infty + A.L^{-3} \qquad (3)$$

which leads to the familiar expression (with L the "cardinal number" of the basis set, typically equal to the highest angular momentum for 1st- and 2nd-row atoms):

$$E_\infty = E(L) + \frac{E(L) - E(L-1)}{\left(\frac{L}{L-1}\right)^3 - 1} \qquad (4)$$

Pansini and Varandas[15,16] proposed a variant where each basis set gets an adjustable, noninteger 'hierarchical number' $\tilde{X}$ associated with it (which for CC methods are 2.71, 3.68, 4.71, and 5.70 for {T,Q,5,6}Z basis sets):

$$E_\infty = E(L) + \frac{E(L) - E(L-1)}{\left(\frac{\tilde{X}_L}{\tilde{X}_{L-1}}\right)^3 - 1} \tag{5}$$

Ranasinghe and Petersson (RP)[17] proposed a two-term shifted formula, the first term of which is actually derived from the CBS pair extrapolation by Petersson and coworkers:[18]

$$E(L) = E_\infty + A.[(L+a)^{-3} + B.(L+a)^{-5}] \tag{6}$$

where, e.g., for the MP2 energy with nZaPa basis sets,[17] $a=1/4$ and $B=-3/2$, and for the (T) energy with the same basis sets, $a=-2/3$ and $B=-7/8$.

Schwenke[19] proposed instead to simply consider a two-point linear extrapolation of the following form:

$$E_\infty = E(L) + A_L[E(L) - E(L-1)] \tag{7}$$

Where $A_L$ is a coefficient specific to the basis set pair and the level of theory. (He recommends eschewing nonlinear 3-point formulas, as they are not size-consistent.)

Of course, the Schwenke formula can be brought into the same form as some of the previous formulas:

$$E_L = E_\infty + \frac{B}{L^\alpha} \quad \text{if} \quad \alpha = \frac{\log\left(1+\frac{1}{A_L}\right)}{\log\left(\frac{L}{L-1}\right)} \tag{8}$$

$$E_L = E_\infty + \frac{C}{(L+1/2)^\alpha} \quad \text{if} \quad \alpha = \frac{\log\left(1+\frac{1}{A_L}\right)}{\log\left(\frac{L+\frac{1}{2}}{L-\frac{1}{2}}\right)} \tag{9}$$

$$E_L = E_\infty + \frac{D}{(L+a)^3} \quad \text{if} \quad a = \frac{1}{\left(1+\frac{1}{A_L}\right)^{1/3} - 1} + 1 - L \tag{10}$$

$$E_L = E_\infty + F.(\tilde{X})^{-3} \quad \text{if} \quad \tilde{X}_L = \tilde{X}_{L-1}\left(1+\frac{1}{A_L}\right)^{1/3} \tag{11}$$

and conversely:

$$A_L = \frac{1}{\left(\frac{L+a}{L-1+a}\right)^\alpha - 1} \tag{12}$$

The RP formula, in particular, has a few desirable features for large $L$: the second term will become negligible, and asymptotically the limiting convergence behavior will be the theoretical[7] $L^{-3}$ behavior of eq. (3).

Table 1 presents Schwenke coefficients $A_L$ for a number of basis set sequences for several levels of theory, together with those obtained from idealized $L^{-3}$ and $L^{-5}$ behavior. In addition, we present the RP shifts $a$ for equivalent RP one-term formulas.

One intriguing feature is that the RP shifts, for large enough basis sets, stay relatively constant. This suggests a convenient way to "extend the range" of an existing extrapolation to larger $L$: by obtaining the RP shift and plugging it into the RP formula together with the next higher value of $L$.

This can, of course, be done in closed form. By writing eq. (10) to extract the shift $a$ in terms of $A_L$ and of $A_{L+1}$, equating the two shifts, and solving for $A_{L+1}$, we obtain for the case of n=3 (singlet-coupled pairs):

$$A_{L+1} = \frac{A_L + 1}{7 + 6A_L \left(1 + \left(1 + \frac{1}{A_L}\right)^{\frac{1}{3}} - 2\left(1 + \frac{1}{A_L}\right)^{\frac{2}{3}}\right)} \quad (13)$$

while for the case of n=5 (triplet-coupled pairs, same-spin correlation energy):

$$B_{L+1} = \frac{B_L + 1}{(80q^4 - 80q^3 + 40q^2 - 10q - 30)B_L - 31} \quad (14)$$

in which $q=(1+1/B_L)^{1/5}$. How well does this work in practice? We can apply this to $A_5$ and see how well the estimated $A_6$ agree with the values obtained by fitting to reference datasets (Table 1, right-hand pane):

TABLE 1. Schwenke coefficients and equivalent Petersson shifts for different basis set pairs {n-1,n}ZaPa or AV{n-1,n}Z

|  | Schwenke coefficients $A_L$, Eq.(7) | | | | Basis sets | Equivalent Petersson shifts $a$, Eq.(6) | | | | Extension of extrapolation | |
| --- | --- | --- | --- | --- | --- | --- | --- | --- | --- | --- | --- |
|  | {6,7} | {5,6} | {4,5} | {3,4} |  | {6,7} | {5,6} | {4,5} | {3,4} | {5,6}→{6,7} | {4,5}→{5,6} |
| $(L+1/2)^{-4}$ (e) | 1.294 | 1.052 | 0.812 | 0.577 | generic | 0.50 | 0.50 | 0.50 | 0.50 | | |
| $L^{-3}$ pure | 1.701 | 1.374 | 1.049 | 0.730 | generic | 0.00 | 0.00 | 0.00 | 0.00 | | |
| $L^{-5}$ pure | 0.861 | 0.672 | 0.487 | 0.311 | generic | 0.00 | 0.00 | 0.00 | 0.00 | | |
| our MP2[a] | 1.852 | 1.503 | 1.127 | | AVnZ | 0.46 | 0.40 | 0.24 | | 1.831 | 1.452 |
| Ref.[17] MP2 | 1.835 | 1.517 | 1.208 | 0.915 | nZaPa | 0.41 | 0.44 | 0.49 | 0.58 | 1.845 | 1.534 |
| ibid. optimized[d] | 1.865 | 1.519 | 1.185 | 0.886 | nZaPa | 0.50 | 0.45 | 0.42 | 0.49 | 1.847 | 1.511 |
| Ref.[10] MP2 | N/A | 1.478 | 1.186 | 0.933 | AVnZ | N/A | 0.32 | 0.42 | 0.64 | 1.805 | 1.512 |
| $(L+0.5)^{-3}$ | 1.865 | 1.537 | 1.211 | 0.889 | generic | 0.50 | 0.50 | 0.50 | 0.50 | | |
| our CCSD[b] | 1.602 | 1.283 | 0.932 | | AVnZ | -0.30 | -0.28 | -0.36 | | 1.609 | 1.255 |
| our CCSD too[b] | 1.605 | 1.232 | 0.917 | | nZaPa | -0.29 | -0.44 | -0.41 | | 1.558 | 1.240 |
| VarandasCCSD | N/A | 1.295 | 0.912 | 0.665 | AVnZ | N/A | -0.24 | -0.43 | -0.21 | 1.621 | 1.235 |
| SchwenkeCCSD | N/A | 1.266 | 0.930 | 0.700 | AVnZ | N/A | -0.33 | -0.37 | -0.09 | 1.592 | 1.253 |
| $(L-0.3)^{-3}$ | 1.602 | 1.276 | 0.953 | 0.636 | generic | –0.30 | –0.30 | –0.30 | –0.30 | | |
| Ditto S pairs | N/A | 1.333 | 1.006 | 0.759 | AVnZ | N/A | -0.12 | -0.13 | 0.09 | 1.660 | 1.330 |
| Ditto T pairs | N/A | 0.755 | 0.530 | 0.454 | AVnZ | N/A | 0.44 | 0.23 | 0.82 | 0.946 | 0.716 |
| Ref.[17] (T) | 1.517 | 1.199 | 0.891 | 0.604 | nZaPa | -0.56 | -0.54 | -0.49 | -0.40 | 1.525 | 1.213 |
| ibid. optimized[d] | 1.580 | 1.164 | 0.849 | 0.600 | nZaPa | -0.37 | -0.64 | -0.62 | -0.41 | 1.490 | 1.171 |
| Schwenke (T) | N/A | 1.248 | 0.810 | 0.730 | AVnZ | N/A | -0.39 | -0.75 | 0.00 | 1.574 | 1.132 |
| our (T)[c] | 1.544 | 1.190 | 0.786 | | AVnZ | -0.48 | -0.56 | -0.82 | | 1.517 | 1.107 |

(a) Present work fitted to MP2-F12/REF-h[10] data obtained using MOLPRO 2015.[20] Aux. basis sets from Ref. [10]
(b) Present work fitted to CCSD-F12 data for 12 closed-shell species in ESI of Ref.[21] at ref. geoms. *ibid.* Original aug-cc-pV7Z basis sets taken from Ref.[13] and refs. therein; updates courtesy of Dr. David Feller (PNNL).
(c) Present work fitted to CCSD(T)/f-limit data at same geoms. obtained following same recipe as Ref.[19] using MOLPRO 2015. Calculations for AV7Z and 7ZaP basis sets performed using GAUSSIAN 09.[22]
(d) Optimized values from Ref.[17], as distinct from Eqs. (11-12) there.
(e) Recently advocated by Feller as a compromise expression[23]

By and large, the 'range-extended' and actual {5,6} extrapolation coefficients agree to 0.05 or better, in many cases to better than 0.02. We can evaluate what a difference of 0.05 in a Schwenke coefficient for, e.g., CCSD/AV{5,6}Z means for the W4-17 dataset[24] of 200 molecules: this works out to just 0.039 kcal/mol RMS. (The exception is Schwenke (T), for which his {4,5} coefficient breaks stride with other evaluations.)

For {6,7} we actually have some values available from the work of RP on the nZaPa sequence: in addition, we have obtained coefficients for the AV{6,7}Z pair by fitting to the CCSD-R12/*spdfghi* total energies of 12 closed-shell species in the supporting information of Tew et al.[21] plus Ne atom.[25] (As a sanity check, we can compare our {5,6} fitted coefficients against these data to those of Hill et al.[10] and of RP for the MP2 case, and to Schwenke and Varandas for the CCSD case: we see that they agree quite well.) Indeed, our 'range-extended' {6,7} coefficients for MP2 and CCSD agree in most cases to 0.02 or better with available optimum values, with an outlier of 0.05 for our reoptimized CCSD/{6,7}ZaPa. For a subset of 186 molecules of W4-17, we were able to perform CCSD/AV7Z calculations as part of an upcoming paper: for that sample, a change by 0.05 in the Schwenke coefficient translates into a change in atomization energies of just 0.018 kcal/mol RMS. Indeed, the difference with pure $L^{-3}$ is only about twice that.

In the context of Pansini and Varandas's extrapolation, the present work favors a 'hierarchical number' for AV7Z and 7ZaPa basis sets of 6.70. This makes the CCSD extrapolations of both the Schwenke and Varandas type functionally equivalent to $E_{corr,CCSD}(\infty) = E_{corr,CCSD}(L) - A/(L-0.30)^3$ for $L \geq 5$.

For MP2, a similarly simple expression works well (Table 1) for $L \geq 4$, $E_2(\infty) = E_2(L) - A/(L+0.50)^3$.

Summarizing, we have a simple and practically workable formula for extending the "reach" of {5,6} basis set extrapolations to larger values of $L$ in general, and a simple closed formula for {6,7}. For the CCSD case, we also point out that our approach is functionally equivalent to Pansini and Varandas, if they extend their "cardinal numbers" one notch further.

*Acknowledgments.* This research was supported by the Israel Science Foundation (grant 1358/15), the Minerva Foundation, the Lise Meitner-Minerva Center for Computational Quantum Chemistry, and the Helen and Martin Kimmel Center for Molecular Design (Weizmann Institute of Science). The author thanks Mr. Nitai Sylvetsky and Profs. Amir Karton and John F. Stanton for helpful discussions, and Dr. David Feller for providing updated machine-readable versions of the AV7Z and (for 2nd-row elements) AV(7+d)Z basis sets. This paper is dedicated to Prof. Viktorya Aviyente (Boğaziçi/Bosphorus University, Istanbul, Turkey) on the occasion of her retirement.